\documentstyle[sprocl,psfig]{article}

\def\be{\begin{equation}}
\def\ee{\end{equation}}
\def\bea{\begin{eqnarray}}
\def\eea{\end{eqnarray}}

\begin{document}

\title{ FERMIONIC AND BOSONIC BUBBLES AND FOAM \footnote{Invited talk given at {\it Collective Excitations
    in Fermi and Bose Systems}, September 14--17, 1998, Serra Negra,
Brazil, to be published by World Scietific, eds. Carlos Bertulani and
Mahir Hussein.} 
}
\author{ \underline{Aurel BULGAC}$^1$,
        Siu A. CHIN$^{2,3}$, Harald A. FORBERT$^2$,
  Piotr MAGIERSKI$^{1,4,5}$ and Yongle YU$^1$ }

\address{ $^1$Department of Physics,  University of Washington,
Seattle, WA 98195--1560, USA  }

\address{ $^2$ Department of Physics and Center for Theoretical Physics,
Texas A\&M University, College Station, TX 77843, USA }

\address{ $^3$ Institute for Nuclear Theory, University of Washington,
Seattle, WA 98195-1550, USA}

\address{$^4$ The Royal Institute of Technology, Physics Department Frescati,
Frescativ\"agen 24, S--10405, Stockholm, SWEDEN }

\address{$^5$ Institute of Physics, Warsaw University of Technology,
ul. Koszykowa 75, PL--00662, Warsaw, POLAND }


\maketitle

\abstracts{ The positioning of one or more bubbles inside a many
fermion or boson system, does not affect the volume, surface or
curvature terms in the liquid drop expansion of the total energy. If
Coulomb effects are irrelevant, the only contribution to the ground
state energy of a system with one or more voids arises from quantum
effects, which is similar to the Casimir effect in vacuum or in
critical phenomena.  We discuss the characteristics of such systems,
the interplay among various effects, such as shell corrections and
chaotic behavior, and briefly mention the role of the temperature and
pairing. 
}

\section{ Introduction }

There are a number of situations when the formation of voids is
favored. When a system of particles has a net charge, the Coulomb
energy can be significantly lowered if a void is created.  The total
energy can be lowered \cite{wheeler,pomorski}, despite an increase in
surface energy.  One can thus naturally expect that the appearance of
bubbles will be favored in relatively heavy nuclei. This situation has
been considered many times over the last 50 years in nuclear physics
and lately similar ideas have been put forward for highly charged
alkali metal clusters \cite{dietrich}.

The formation of gas bubbles is another suggested mechanism which
could lead to void(s) formation \cite{moretto}. The filling of a
bubble with gas prevents it from collapsing.  In a similar manner the
air bubbles in an ordinary glass of water or steam bubbles in boiling
water are stabilized.  Various heterogeneous atomic clusters
\cite{saito} and halo nuclei \cite{austin} can be thought of as some
kind of bubbles as well.  In these cases, the Fermions reside in a
rather unusual mean--field, with a very deep well near the center of
the system and a very shallow and extended one at its periphery.
Since the amplitude of the wave function in the semiclassical limit is
proportional to the inverse square root of the local momentum, the
single particle wave functions for the weakly bound states will have a
small amplitude over the deep well. If the two wells have greatly
different depths, the deep well will act almost like a hard wall (in
most situations).

Several aspects of the physics of bubbles in Fermi and Bose systems
have not been considered so far in the literature.  It is tacitly
assumed that a bubble position has to be determined according to
symmetry considerations. For a Bose system one can easily show that a
bubble has to be off--center \cite{chin}. A system of independent and
noninteracting Bosons is a particularly simple system at $T=0$, since
all the particles will have the same single particle wave function. If
now one would like to ``drill'' a hole, it would be more costly to do
that in a region where the amplitude of the wave function is large,
namely in the center of the system. A hole in a region where the wave
function has a node should be almost ``painless''. The greater the
amplitude of the wave function the greater the ``pain'', and thus
energetically more costly. In a Bose system a hole could be created by
injecting a large fullerene \cite{chin}.

In the case of a Fermi system the most favorable arrangement is not
obvious and the determination of its characteristics is nontrivial
\cite{aurel}. The energy of a many fermion system has the general form
\begin{equation}
E(N)=E_{LD}(N)+E_{sc}(N) = e_vN +e_sN^{2/3}+e_cN^{1/3} + E_{sc}(N),
\label{eq:liq}
\end{equation}
where $E_{LD}$ is the smooth liquid drop part of the total energy and
$E_{sc}$ is the pure quantum shell correction contribution to the
total energy. We shall consider in this work only one type of Fermions
with no electric charge. In a nuclear system the Coulomb energy
depends rather strongly on the actual position of the bubble. In an
alkali metal cluster, as the excess charge is always localized on the
surface, the Coulomb energy is essentially independent of the bubble
position.

Once a bubble is formed, its position does not affect the volume,
surface or curvature terms in the liquid drop mass formula. Only
quantum effects are left responsible for determining the optimal
geometrical configuration.  The character of the shell corrections is
strongly correlated with the existence of regular and/or chaotic
motion \cite{balian}. If a spherical bubble appears in a spherical
system and if the bubble is positioned at the center, then for certain
``magic'' Fermion numbers the shell correction energy $E_{sc}$, and
hence the total energy $E(N)$, has a very deep minimum. However, if
the number of particles is not ``magic'', in order to become more
stable the system will in general tend to deform.  Real deformations
lead to an increased surface area and liquid drop energy.  On the
other hand, merely shifting a bubble off-center deforms neither the
bubble nor the external surface and therefore, the liquid drop part of
the total energy of the system remains unchanged.

Moving the bubble off--center can often lead to a greater stability of
the system due to shell correction energy effects.  In large systems
the relative importance of the shell correction energy $E_{sc}$ is
small -- it increases with $N$ at a rate even smaller than the
curvature energy.

In recent years it was shown however that in a 2--dimensional annular
billiard, i.e. in the 2--dimensional analog of spherical bubble
nuclei, the motion becomes more chaotic as the bubble is moved further
from the center \cite{bohigas}.  This effect is expected to diminish
the importance of the shell corrections and thus raise again, the
question of whether displacing the bubble off--center can lead to more
stable configurations.

One can anticipate that the relative role of various periodic orbits
(diameter, triangle, square etc.) is modified in unusual ways in
systems with bubbles.  In 3--dimensional systems the triangle and
square orbits determine the main shell structure and produce the
beautiful supershell phenomenon \cite{balian,ben}.  A small bubble
near the center will affect only diameter orbits.  After being
displaced sufficiently far from the center, the bubble will first
touch and destroy the triangle orbits.  In a 3--dimensional system
only a smaller fraction of these orbits will be destroyed.  Thus one
expects that the existence of supershells will not be critically
affected, but that the supershell minimum will be less pronounced.  A
larger bubble will simultaneously affect triangular and square orbits,
and thus can have a dramatic impact on both shell and supershell
structure.

It is natural to consider also the formation of two or more bubbles at
the same time, in finite, infinite or semi--infinite systems.  For a
sufficiently large bubble density a new form of matter can be created,
foam. One might argue that sometimes a ``misty'' state could be more
likely. As in the case of percolation, whether a ``foamy'' or a
``misty'' state would be formed, should strongly depend on the average
matter density. At very low average densities, formation of droplets
is more likely, while at higher average densities (lower than the
equilibrium density however) the formation of a foam is more
probable. Similar ideas have been explored for the case of neutron
matter and various ``exotic'' structures have been found, such as
bubbles, plates/lasagna, and rods/spagetti \cite{chris}.  The
energetics of two or more bubbles, their relative placements and
positions with respect to boundaries, their collisions and bound state
formation, their impact on the role played by periodic or chaotic
trajectories, and their temperature dependence, are but a few in a
long list of challenging questions.  A plethora of new, extremely soft
collective modes is thus generated.  The character of the response of
such systems to various external fields is an extremely intricate
issue.  Since the energy of the system changes only very little while
the bubble(s) is being moved, a slight change in energy can result in
large scale bubble motion.  Such a system may prove to be an extremely
sensitive ``detector''.

 In Section 2 we describe two numerical methods we have developed in
order to determine the single--particle spectrum of quantum billiards.
In Section 3 we describe the case of one bubble in a spherical system,
in Section 4 the case of two bubbles in an infinite medium and we
conclude with several final remarks in the last Section.

\section{ Methods to determine the single--particle spectrum for
quantum billiards of arbitrary shapes }

The change of the total energy of a many Fermion system can be
computed quite accurately using the shell corrections method, once the
single--particle spectrum is known as a functions of the shape of the
system \cite{strut}.  We describe here briefly two methods we have
developed in order to determine the eigenvalue spectrum of quantum
billiards of arbitrary shapes: the conformal mapping method and the
boundary overlap method (BOM). The methods known in literatures are
suited to so called star shaped domains \cite{saraceno} and they meet
with significant and/or unsurmountable difficulties when the domains of
interest have unusual topologies and/or shapes.

\subsection{ Conformal mapping of the  eigenvalue problem }

This method is especially suited for treating systems with round
bubbles.  By means of an appropriately chosen conformal
transformation, a single--connected 2--dimensional region can be
transformed into a circular annulus. We shall consider here three
different cases: {\it i}) a circular hole inside a circular cavity;
{\it ii}) two circular holes in an infinite medium, and {\it iii}) a
hole in a semi--infinite medium.

Even though the partial differential equation is the same:
\begin{equation}
-[\partial ^2_u + \partial ^2_v ]\Psi (u,v) = k^2 \Psi (u,v),
\end{equation}
the Dirichlet boundary conditions in the first two cases are imposed
on the following two circles
\begin{equation}
\Psi (u,v)|_{\cal{B}} =0  \quad \mathrm{where} \quad
{\cal{B}} = \{ u^2+v^2 =R^2=1 \cup (u-d)^2+v^2=a^2 \} ,
\end{equation}
(we have chosen here a coordinate scale such that $R=1$) $0\le a\le
R$, $0\le d\le R=1$ and $0\le a+d\le R=1$ in the first case and $0\le
a$ and $a+R=a+1\le d$ in the second case. By means of the conformal
transformation
\begin{eqnarray}
w&=&\frac{z-c}{1-zc}, \quad w=u+iv, \quad z=x+iy\\
c&=& \frac{1+d^2-a^2-\sqrt{(1+d^2-a^2)^2-4d^2}}{2d}
\end{eqnarray}
the unit circle $u^2+v^2 =R^2=1$ is mapped onto itself, while the
circle $(u-d)^2+v^2=a^2$ becomes a concentric circle with a radius:
\begin{equation}
r= \left | \frac{d+a-c}{1-c(d+a)} \right |
\end{equation}
with $r< 1$ in the first case and $r> 1$ in the second case.  The
third case, when the boundary is
\begin{equation}
{\cal{B}} = \{(u+d)^2+v^2=a^2\cup \mathrm{Re} \; z =u= 0 \}
\end{equation}
can be handled with a similar conformal transformation, namely
\begin{equation}
w=\frac{A+1}{A-1} \; \frac{z+A(d+a)}{z-A(d+a)}, \quad
A=\sqrt{\frac{d-a}{d+a}}.
\end{equation}
Under this transformation the circle in the left plane (by convention
$d>0$) becomes the unit circle with the center at the origin, while
the imaginary axis becomes a concentric circle of radius
\begin{equation}
r= \frac{1+A}{1-A}\ge 1 .
\end{equation}
The partial differential equation then becomes
\begin{equation}
-J(x,y) [\partial ^2_x + \partial ^2_y ]\Psi (x,y) = k^2 \Psi (x,y),
\end{equation}
where $x=\rho \cos \phi$, $y=\rho \sin \phi$ and where for the first
two geometries
\begin{equation}
J(x,y) = \left |\frac{\partial z}{\partial w}\right |^2 =
\frac{[(1-cx)^2+c^2y^2]^2}{(1-c^2)^2} =
\frac{[1+c^2\rho ^2-2c\rho \cos \phi ]^2}{(1-c^2)^2}
\end{equation}
and
\begin{equation}
J(x,y) =
\frac{(1+A)^2[(x-A)^2+y^2]^2}{4A^2(1-A)^2} =
\frac{(1+A)^2[\rho ^2 + A^2-2A\rho \cos \phi ]^2}{4A^2(1-A)^2}
\end{equation}
in the case of the third geometry.

One can now expand the eigenfunction $\Psi (x,y)$ in the complete set
of states of the on--center problem with $J(x,y)=1$,
\begin{eqnarray}
& &\Phi _{n,m}(\rho, \phi)= C_n
[    J_m(p_n\rho ) Y_m(p_n r ) - Y_m(p_n\rho ) J_m( p_nr )]\exp (im\phi ),\\
& &[ J_m(p_nR )    Y_m(p_n r ) - Y_m(p_nR )    J_m( p_nr )]=0, \\
& & \langle \Phi _{nm}|\Phi _{n'm'}\rangle = \delta _{nn'}\delta
_{mm'}
\end{eqnarray}
where $m$ is the magnetic quantum number, $J_m(x)$ and $Y_m(x)$ are
the cylindrical Bessel functions of the first and second kind, $C_n$
is a normalization constant, $p_n^2$ are the corresponding eigenvalues
of the on--center problem. Using the following representation of the
eigenfunction $\Psi (\rho , \phi )$
\begin{equation}
\Psi (\rho , \phi )=\sum _{nm} A_{nm}\frac{1}{k_n}\Phi _{n,m}(\rho, \phi)
\end{equation}
one can easily establish that the expansion coefficients $A_{nm}$ for
the eigenfunctions and the eigenvalues of the off--center problem can
be determined from the following eigenvalue problem
\begin{equation}
\sum _{n'm'}
k_n\langle \Phi _{nm}| J|\Phi _{n'm'}\rangle k_{n'} \;
 A_{n'm'} = k^2  A_{nm}.
\end{equation}
This matrix is block--diagonal as the matrix elements $\langle \Phi
_{nm}| J|\Phi _{n'm'}\rangle $ vanish for $|m-m'| > 2$.

The 3--dimensional case, when circles become spheres and lines become
planes, can be easily treated in a similar manner. For all the
geometries corresponding to the 2--dimensional cases described above
the problem has an axial symmetry, corresponding to a simple
reflection symmetry $y \rightarrow -y$ in the 2--dimensional
problem. It is easy to show that the eigenfunctions have the following
form
\begin{equation}
\Psi (\rho, \phi , z) =
\frac{\psi _m (\rho , z)\exp ( im\phi )}{\sqrt{\rho }}
\end{equation}
where $\Psi (\rho , z)$ satisfy the partial differential equation
\begin{equation}
\left [ -(\partial ^2 _\rho + \partial ^2 _z) +
\left ( m^2 -\frac{1}{4}\right ) \frac{1}{\rho ^2} \right ]
\psi _m (\rho , z)   = k^2 \psi _m (\rho , z), \label{eq:hem3}
\end{equation}
with the obvious boundary condition $\psi ( \rho =0, z) =0$. By
performing similar conformal transformations in the upper
half--complex plane $z+i\rho$, one can reduce the 3--dimensional
problems to one similar to the 2--dimensional cases discussed above.

\subsection{ Boundary overlap method }

The solution of the Helmholtz equation for the 2--dimensional annular
billiard can be represented as follows
\bea
\Psi (u,v)&=& \sum _{m} C_m \Phi _m(k\rho , \phi ),\\
\Phi _m(k\rho , \phi ) &=&
[J_m(k\rho )Y_m(ka)-J_m(ka)Y_m(k\rho )]\exp (im\phi ),
\eea
with the yet undetermined coefficients $C_m$. Obviously, this basis
set is not unique, but it is the best suited one for the problem at
hand \cite{aurel}.  By requiring that the boundary overlap vanishes
\bea
& &\frac{1}{2\pi R}\oint _{\cal{B}} dl
\sum _{m_1,m_2} C_{m_1}^* \Phi _{m_1}^*(k_0 \rho ,\phi )
\Phi _{m_2}(k_0 \rho ,\phi )  C_{m_2} \\
&=& \sum _{m_1,m_2} C_{m_1}^*
{\cal{O}}_{m_1,m_2}(k_0)C_{m_2} =0 \label{eq:quant}
\eea
one can determine the eigenvalues and the eigenvectors. This
quantization condition is satisfied only for discrete values of $k_0$.
For arbitrary values of $k$ one can introduce the eigenvalues and
eigenvectors of the boundary overlap matrix (BOM) ${\cal{O}}(k)$
\be
{\cal{O}}(k)C_\alpha = \lambda _\alpha (k)C_\alpha .
\ee
>From the non--negativity of the boundary norm it follows that for real
values of $k$ these eigenvalues satisfy the inequality $\lambda
_\alpha (k)\ge 0$ and only for an eigenvalue of our initial problem
$\lambda _\alpha (k_0)= 0$. In the neighborhood of such an eigenvalue
$\lambda _\alpha (k) \propto (k-k_0)^2 + \dots $.The basic idea behind
our approach is to analytically continue BOM ${\cal{O}}(k)$ into the
complex $k$--plane and to compute around a contour ${\cal{C}}$ the
integral
\be
{\cal{N}}({\cal{C}})=\oint _{\cal{C}} \frac{dk}{4\pi i}
\sum _\alpha \frac{\lambda _\alpha ^\prime (k)}{\lambda _\alpha (k)},
\ee
where $\lambda _\alpha ^\prime (k)$ is the derivative of the
eigenvalue $\lambda _\alpha (k)$ with respect to $k$, which can be
evaluated using the simple formula
\be
\lambda _\alpha ^\prime (k) =
C_\alpha^\dagger (k) {\cal{O}}^\prime (k) C_\alpha (k).
\ee
Above, ${\cal{O}}^\prime (k)$ is the derivative of $ {\cal{O}} (k)$
with respect to $k$ and the eigenvectors are normalized as usual,
$C_\alpha^\dagger (k) C_\alpha (k) =1$ (vector and matrix
multiplication rules are implied in previous formulae).  Thus
${\cal{N}}({\cal{C}})$ is equal to the number of eigenvalues (counting
the degeneracies as well) located on the segment of the real $k$--axis
enclosed by the contour ${\cal{C}}$. Using a similar formula one can
compute the energy of an ${\bf N}$--Fermion system
\be
E_{\bf N}= \oint _{\cal{C}} \frac{dk}{4\pi i}
\sum _\alpha \frac{\lambda _\alpha ^\prime (k)}{\lambda _\alpha (k)} k^2,
\ee
where the number of Fermions is given by
\be
{\bf N}={\cal{N}}({\cal{C}}).
\ee
We shall not dwell here on various numerical subtleties, as the issues
regarding the numerical implementation of the BOM method will be
described in more detail elsewhere \cite{aurel}.

\section{ One bubble in a finite system }

The simplest case to consider is a circular 2--dimensional Fermi or
Bose system, in which one can ``drill'' a circular hole and determine
how the ground state energy of the system changes when the hole is
displaced from the center towards the edge of the system.

In Fig. 1 we show the unfolded single--particle spectrum for the case
of a bubble of half the radius of the system, $a=R/2$, as a function
of the displacement $d$ of the bubble from the center. The unfolded
single--particle spectrum is determined from the Weyl formula
\cite{weyl} for the average cumulative number of states (for spinless
particles) with energy less than $k^2$ as follows
\bea
N_0(k) &=& \frac{S}{4\pi}k^2-{\frac{L}{4\pi}k+\frac{1}{12\pi}\oint
_{{\cal{B}}}}dl\kappa (l) = \frac{R^2-a^2}{4}k^2-\frac{R+a}{2}k,\\
e_n &=& N_0(k_n).
\eea
In the above formula, $S$ is the area of the 2--dimensional system,
$L$ is its perimeter, $\kappa (l) $ is the local average curvature
along the perimeter and $l$ is the length coordinate, thus $L=\oint
_{{\cal{B}}}dl$, and $k_n^2$ is the actual $n$--th energy eigenvalue.
By construction the unfolded spectrum $e_n$ has an unit average level
density.  When the bubble is at the center, the problem is
rotationally symmetric with a highly (quasi)degenerate
single--particle spectrum.  The existence of symmetries, which give
rises to high degeneracies in the single particle spectrum, is the
basic reason for the existence of ``magic numbers'' and extra stable
nuclei and atoms.  If one were to construct a nearest--neighbor level
splitting distribution, a distribution very different from the Wigner
surmise should emerge in this case (we shall not display it here,
however). Typically for integrable systems the nearest--neighbor level
splitting distributions have a Poissonian character.  As the bubble is
moved off center, the classical problem becomes more chaotic
\cite{bohigas}. One can expect that the single particle spectrum would
approach that of a random Hamiltonian \cite{gian} and that the
nearest--neighbor splitting distribution would be given by the Wigner
surmise\cite{mehta}.  The spectrum for chaotic systems is typically
very ``rigid'', it shows a relatively low level of fluctuations among
distant level. This particular and remarkable property of the spectrum
is best illustrated by evaluating the so called $\Delta _3$ statistics
\cite{mehta}. The $\Delta _3$ statistics is the most reliable quantity
used to establish whether a given quantum spectrum is regular or
chaotic, depending on whether this statistic is closer to the Poisson
or to the GOE/GUE/GSE limits (here GOE/GUE/GSE represent one of the
universal Gaussian ensembles of random Hamiltonians \cite{mehta}).  A
random Hamiltonian would imply that ``magic'' particle numbers are as
a rule absent.  There is a large number of level crossings in Fig. 1,
but in spite of that one can definitely see a significant number of
relatively large gaps in the spectrum.  If the particle number is such
that the Fermi level is at a relatively large gap, one can expect that
the system at the corresponding ``deformation'' is very stable. This
situation is very similar to the celebrated Jahn--Teller effect in
molecules.  A simple inspection of Fig. 1 suggests that for various
particle numbers the energetically most favorable configuration can
either have the bubble on-- or off--center.  Consequently, a ``magic''
particle number could correspond to a ``deformed'' system. In this
respect this situation is a bit surprising, but not unique. It is well
known that many nuclei prefer to be deformed, and there are
particularly stable deformed ``magic'' nuclei or clusters
\cite{strut,bohr,lew}.

For a (noninteracting) many Boson system only the lowest single
particle level would be relevant. As it was shown by two of us
\cite{chin}, it is energetically favorable to expel the bubble, as it
is more costly to ``drill'' a hole in the center, where the
single--particle wave function has a maximum, than at the edge of the
system.

The variation of the ground state energy of an interacting
$N$--Fermion system, with respect to shape deformation or other
parameters, is accurately given by the shell correction energy
\cite{strut}
\bea
\delta E(N) &= &\sum _{n=1}^N k_n^2 - \int _0 ^{k_0} dk
\frac{N_0(k)}{dk}k^2,\label{eq:sc} \\
N &=& N_0(k_0). 
\eea
In our case, the eigenspectrum and the shell correction energy are
functions of $N$, $R$, $a$ and $d$.  When the particle number $N$ is
varied at constant density, we have $R=R_0N^{1/2}$ and $a=a_0N^{1/2}$
in 2--D and $R=R_0N^{1/3}$ and $a=a_0N^{1/3}$ in 3--D.  There is a
striking formal analogy between the energy shell correction formula
and the recipe for extracting the renormalized vacuum Casimir energy
in quantum field theory \cite{cas} or the critical Casimir energy in a
binary liquid mixture near the critical demixing point \cite{fish}.
In computing the shell correction energy the ``trivial'' liquid drop
or the macroscopic dependency of the total energy, see
Eq. \ref{eq:liq}, on the geometry of the system cancels out and only
the pure quantum effects remain. In the $N\rightarrow \infty $ limit
the shell correction energy becomes irrelevant.  For a very large
3--dimensional Fermionic systems, it was shown by Strutinsky and
Magner \cite{strutin} that $\delta E(N) \propto N^{1/6}$, which is
significantly less than even the curvature corrections, which behave
as $\propto N^{1/3}$.  In 2--dimensions the curvature term in the
energy is $N$--independent.  In Fig. 2 we show the contour plot of the
shell correction energy for the same system with the (unfolded)
single--particle spectrum shown in Fig. 1 ($a=R/2$) as a function of
the bubble displacement $d$ versus $N^{1/2}$. The overall regularity
of ``mountain ridges' and ``canyons'' is at least somewhat unexpected.
One can see that various mountain tops and valleys form an alternating
network almost orthogonal to the ``mountain ridges'' and
``troughs''. For some $N$'s the bubble ``prefers'' to be in the
center, while for other values that is the worst energy
configuration. For a given particle number $N$ the energy is an
oscillating function of the displacement $d$ and many configurations
at different $d$ value have similar energies.

A bubble with a radius $a=R/2$ is quite large and when it is at the
center, it it tangent to all classical triangular orbits.  When the
bubble is displaced off center one naturally expect that there are no
remnants of these orbits and therefore no supershell phenomenon in
this two dimensional system.  It is therefore surprising that even for
relatively large displacements the contour plot of the shell
correction energy retained significant structure, somewhat at odds
with the naive expectation that the role of triangular and square
orbits should be suppressed.

What is changed if the bubble has a smaller radius? In Fig. 3 we show
the unfolded single--particle spectra for a bubble with $a=R/5$. When
compared to the spectra in Fig. 1 one can see that the number of level
crossing is significantly smaller.  One can also show that in the
limit of a vanishing bubble radius ``nothing happens'', see
e.g. Ref. \cite{chin}. As a result, the shell correction energy
contour plot has less structure.  This is shown in Fig. 4.  The
amplitude of the fluctuations are now smaller and the energy changes
less in the $d$--direction. Thus a system with a smaller bubble is
significantly softer.

Simply for the lack of space we shall not present here any results for
the 3--dimensional case. It suffices to mention that the overall
qualitative picture is similar to the 2--dimensional case.

\section{ Two identical bubbles in an infinite medium }

By applying similar techniques one can study the case of two identical
bubbles in an infinite medium (or a single bubble near the boundary of
a semi--infinite medium).  The analysis we have performed in this case
is less complete.  Although the single--particle spectrum is
continuous, our discretization procedure produces only a discrete
spectrum.  We can show however, that the ``wave functions'' so
determined are localized mostly between the two bubbles.  The rest of
the wave functions, which we do not determine, are in some sense ``far
away'', and are not very sensitive to the relative positions of the
two bubbles. Increasing the size of the numerical basis set leads to a
denser eigenvalue spectrum, but the dependence of the single--particle
eigenvalues on the separation between the two bubbles remains
unchanged. We have verified this by performing calculations with
increasingly larger basis sets, up to four thousand basis states.  In
Fig. 5 we display the energy eigenvalues multiplied with $d^\alpha$,
i.e. $k_n^2(d)d^\alpha$, where $d$ is the minimum distance between the
two bubble surfaces.  As one can see, for relatively small, but not
too small separations, the single--particle spectrum can be fairly
described by a simple power law, $ k^2_n(d)\propto \varepsilon
_n(d)/d^{\alpha}$, where $\varepsilon _n(d)$ is a slowly varying function
of $d$ and $\alpha \approx 2/5$. This naturally implies that the total
energy of such a system has a similar distance dependence and thus two
bubble in an infinite medium will repel each other with a simple power
law potential $U(d) \propto d^{-\alpha }$. This result has a limited
range of validity and cannot be expected to be correct at larger
separations.  At large distances one can show that the bubble--bubble
interaction is similar to the Ruderman--Kittel interaction in
condensed matter physics, i.e. an oscillatory potential whose
amplitude decreases as a power law.  When the bubble--bubble
separation is much larger than the bubble radii and the Fermi
wavelength one can estimate the bubble--bubble interaction using the
linear response theory. Let us first assume that the two bubbles can
be described as two small impurities separated by a distance ${\bf
d}$. The interaction energy is given by
\be
E_{12}({\bf d}) = \int d{\bf r}_1\int d{\bf r}_2
V_1({\bf r}_1 ) \chi ({\bf r}_1- {\bf r}_2 - {\bf d})V_2({\bf r}_2),
\ee
where $\chi ({\bf r}_1 -{\bf r}_2 -{\bf d})$ is the static form factor
or the Lindhard response function of a homogeneous Fermi gas and
$V_1({\bf r}_1 )$ and $V_2({\bf r}_2)$ are the potentials describing
the interaction of these two impurities with the Fermion
background. Since the bubbles cannot be considered as weak impurities
the potentials should be replaced by the corresponding scattering
amplitudes, namely
\be
T_{1,2} = V_{1,2}+V_{1,2}GT_{1,2},
\ee
where $G$ is the single--particle propagator.  The formula for the
energy shift becomes somewhat more complicated, however, the
dependence on the bubble--bubble separation is not drastically
modified. The asymptotic behavior is
\be
E_{12}({\bf d}) \propto \frac{\cos (2 k_Fd)}{d^3},
\ee
where $k_F$ is the Fermi momentum.  At very small separations on the
other hand, as our numerical result seem to imply, the repulsion could
become stronger than a simple $\propto d^{-\alpha }$ potential, where
$\alpha$ was determined above, see Fig. 5. When two bubbles are at a
relative distance much smaller than the bubble radii, one can replace
the two bubbles with two infinite parallel plates. It is
straightforward to show that the interaction energy in this case is a
repulsive power law potential $U(d)\propto d^{-\alpha}$, where $\alpha
=2$ or $\alpha =1$ depending on whether the particle number or the
particle density is kept constant while the distance between the
plates is varied.  This result thus supports qualitatively the above
inference.

The short distance repulsion will prevent two bubbles from collapsing
into a single bigger bubble, even though this will lower the surface
energy. One could expect however the formation of ``bubble molecules''
of various sizes.

\section{ Conclusions}

We did not have here the space to discuss in any detail the influence
of temperature. Naively since one expects rising temperature to smooth
out shell effects, the same should happen to systems with bubbles.
Thus at finite temperature, the relative position of a bubble inside a
many--body system or the relative positioning of two bubbles in an
infinite medium, should be almost insensitive to the bubble--bubble
separation.  What is wrong with this type of argument however is the
fact that at finite temperatures one should instead consider the free
energy. The entropy of the system increases as one displaces the
bubble off-center, due to a contribution, which can be called
positional entropy, $S({\bf d}) = \ln d +{\rm const}$.  Moreover, making
more bubbles could lead to a further decrease of the free energy,
even though the energy might increase. Thus the problem of one or more
bubbles at finite temperatures has its own special intricacies.

Pairing correlations can lead to a further softening of the potential
energy surface of a system with one or more bubbles. We have seen 
the energy of a system with a single bubble is an oscillating function
of the bubble displacement.  When the energy of the
system as a function of this displacement has a minimum, the Fermi
level is in a relatively large gap where the single--particle level
density is very low. When the energy has a maximum, just the opposite
is true.  On the other hand pairing correlations will be significant
when the Fermi level occurs in a region of high single--particle level
density. It is thus natural to expect that the total energy is lowered
by paring correlations at ``mountain tops'', and be less affected at
``deep valleys", which ultimately leads to further leveling of the
potential energy surface.

We also did not study other types of boundary conditions, Neumann or
mixed boundary conditions, at the bubble boundary and/or at the system
boundary. Depending on the nature of the bubble, it may be necessary
to consider these cases in the future.  The change in boundary
conditions can lead to completely different conclusions in specific
situations.  

We can make a somewhat weaker argument for why only
consider spherical bubbles.  In the process of deforming a bubble the
surface energy changes. However, for very large particle numbers
the shell correction energy is parametrically much smaller than the
surface energy. Thus, at least for large systems, spherical bubbles
should be energetically favored (if the surface tension is positive).

A system with one or more bubbles should be a very soft system.  The
energy to move a bubble is parametrically much smaller than any other
collective mode. For this reasons, once a system with bubbles is
formed, it could serve as an extremely sensitive ``measuring device'',
because a weak external field can then easily perturb the bubble(s)
and produce a system with a completely different geometry.  In
Ref. 7 a Bose--Einstein condensate detector was suggested,
which however is a detector of a slightly different nature than the
one we are suggesting here. There the Bose character of the particles
lead to an enhancement of the interaction between the bubble and the
condensate.

There are many systems where one can expect that the formation of
bubbles is possible. Known nuclei are certainly too small and it is
difficult at this time to envision a way to create nuclei as big as
those predicted in Ref. 2. On the other hand voids can
be easily conceived to exist in neutron stars \cite{chris}. Metallic
clusters with bubbles are much easier to imagine. Another realistic
system to consider is two or more fullerenes in either liquid sodium
or liquid mercury.  In the case of sodium the electron wave functions have a
node at the fullerene surface while in the case of mercury they do not. Since
fullerenes do not melt easily, one can also consider other liquid
metals.  There is another experiment that one can suggest, to place a
metallic ball inside a superconducting microwave resonator of the type
studied in Ref. 23 and study the ball energetics and maybe
even dynamics.

\section*{Acknowledgments}

AB thanks B.Z. Spivak for discussions concerning the
impurity--impurity interaction, D. Tom\'anek for suggesting a possible
experiment with two fullerenes in a liquid metal, C.H. Lewenkopf for
several literature leads concerning existing numerical methods for
quantum billiards and G.F. Bertsch for suggesting a simple
1--dimensional ``billiard'' to test our methods.  SAC would like to
thank the Institute for Nuclear Theory at the University of Washington
for hospitality during the Atomic Clusters Program, Summer, 1998. PM
thanks the Nuclear Theory Group in the Department of Physics at the
University of Washington for hosting his stay. AB and YY acknowledge
the DOE financial support and SAC and HAF acknowledge NSF financial
support.

\section*{ Figure captions }

\begin{itemize}

\item Fig. 1. The unfolded single--particle spectrum $e_n(d)$, for a
circular cavity of unit radius $R=1$ with a bubble of radius $a=R/2$,
as a function of the displacement of the bubble from the center of the
cavity $d$. The even-- and odd--parity states are shown separately, in
the left and right subplots respectively. The parity is defined with
respect to the reflection with respect to the line joining the bubble center
to the center of the cavity.

\item The contour plot of the shell correction energy $\delta E(N)$, 
see Eqs. \ref{eq:sc}, as a function of the $N^{1/2}$ and $d$.  The
radius of the system was chosen for each particle number as
$R(N)=R_0N^{1/2}= N^{1/2}$ and in the figure $d$ is the actually the fraction
$d/R(N)$. The size of the bubble also scales with the particle number
as $a(N)=R(N)/2$.

\item Fig. 3. The same as in Fig. 1 but for a smaller bubble with the
radius $a=R/5$.

\item Fig. 4 The same as in Fig. 2 but for for a smaller bubble with
the radius $a(N)=R(N)/5$.

\item Fig. 5 The scaled single--particle spectrum $k_n^2(d)d^\alpha$,
for two bubbles of unit radius $R=1$, separated by a distance $d$. The
exponent was determined numerically to be $\alpha \approx 0.425$.

\end{itemize}

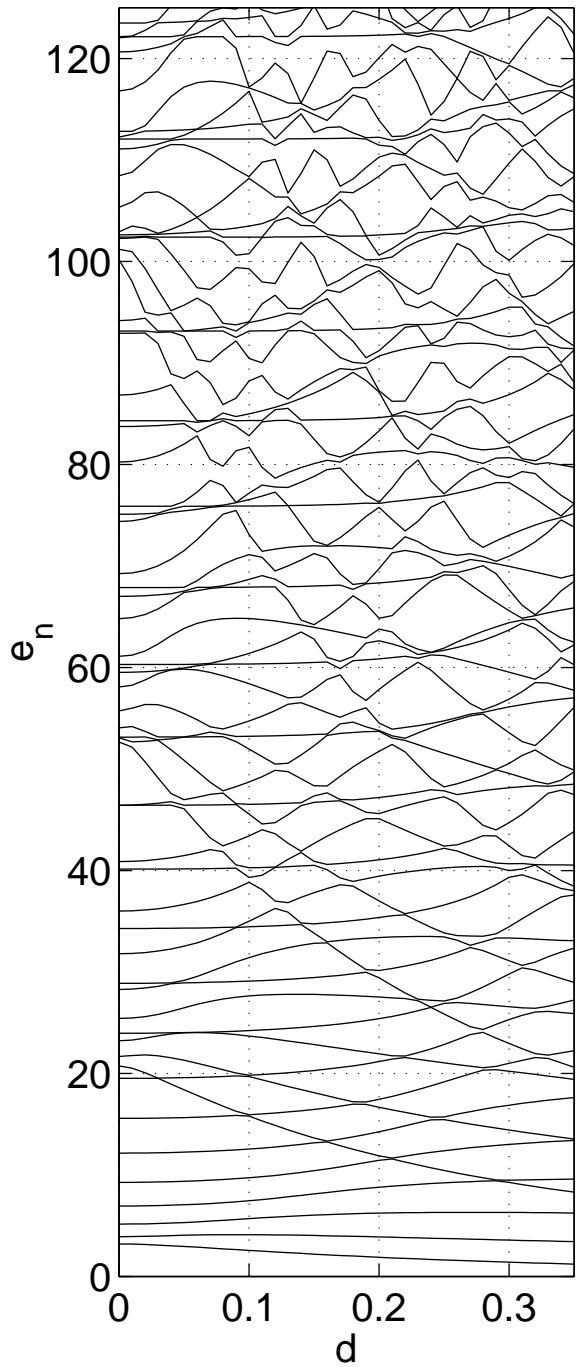
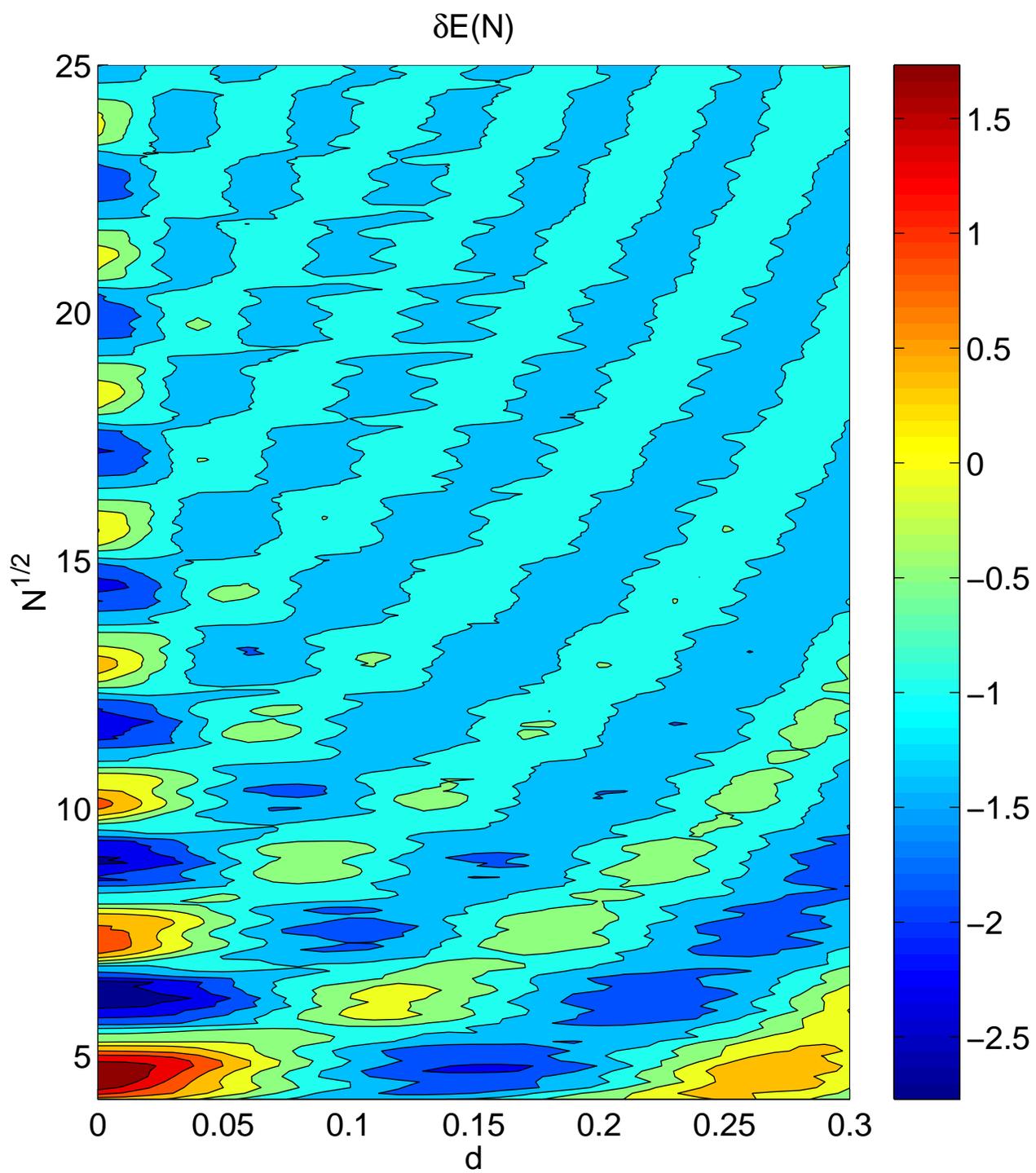
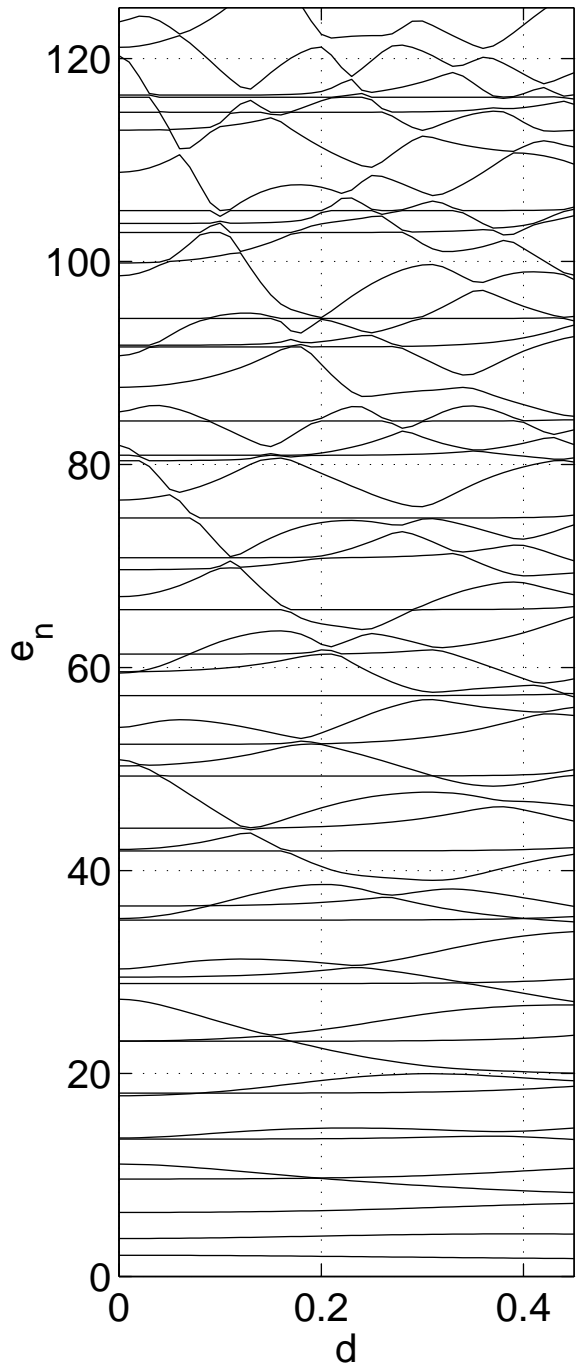
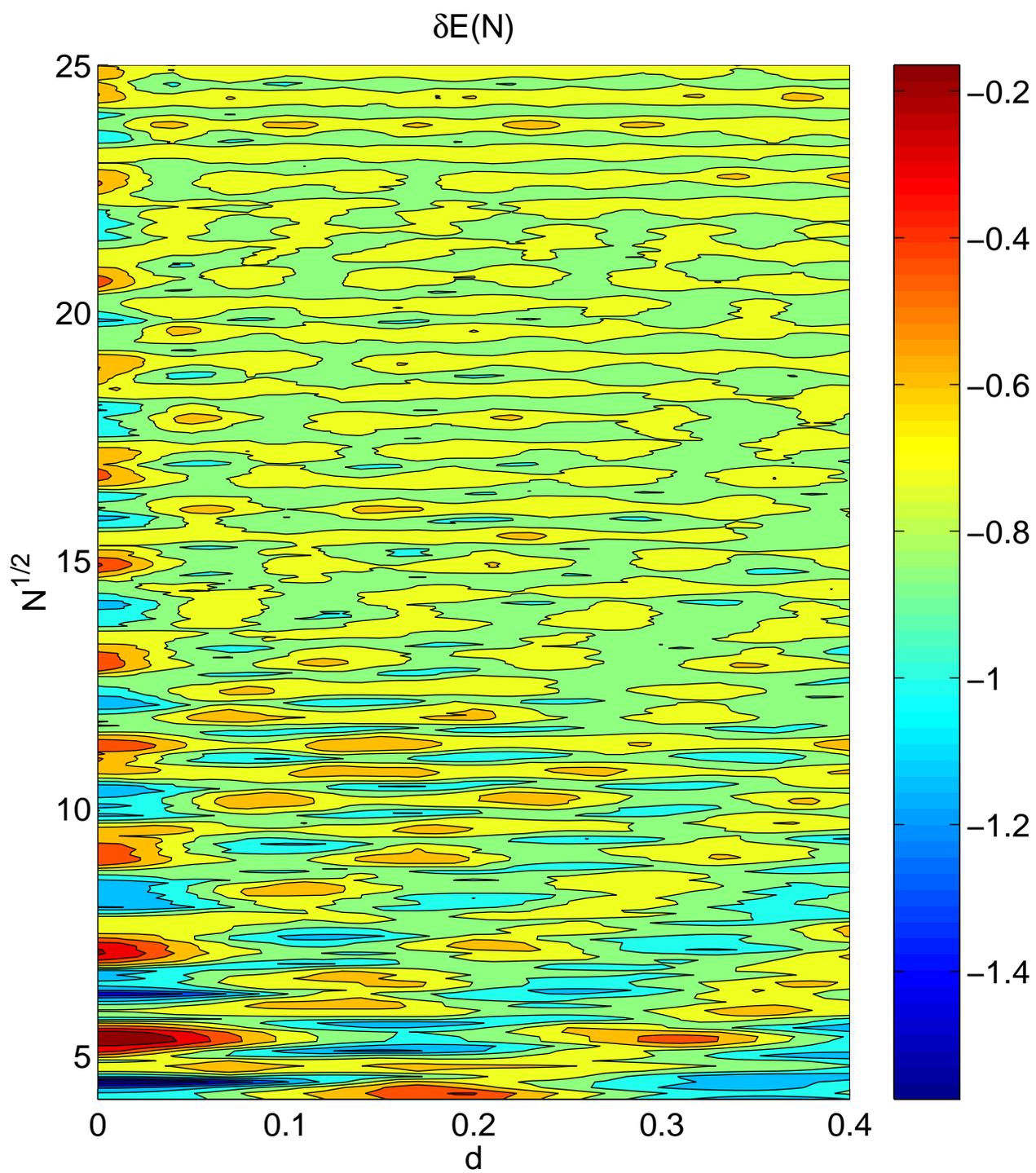
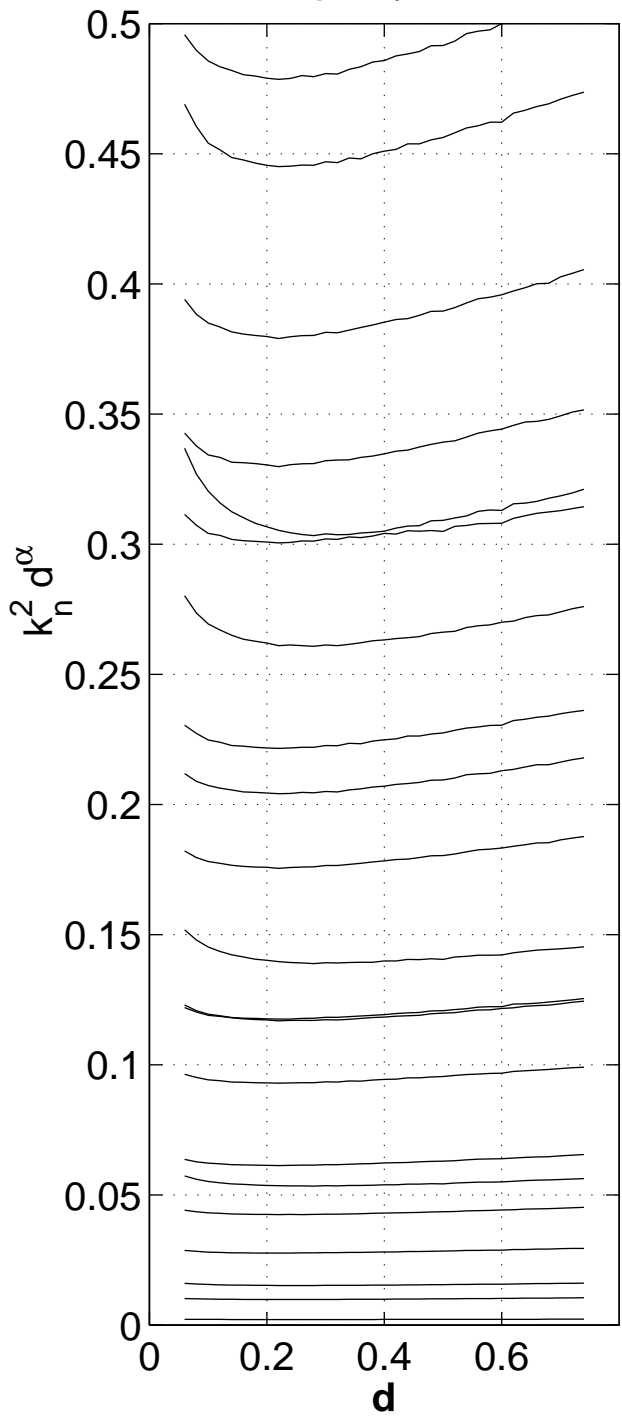
\end{document}